%Paper: gr-qc/9407035
%From: Eric Benedict <eben@buphy.bu.edu>
%Date: Fri, 22 Jul 1994 19:05:53 GMT

%manuscript written in plain TeX
%%%
\input phyzzx

\let\refmark=\PLrefmark

\def\Lga{{\cal L}_{\rm gauge}}
\def\Lge{{\cal L}_{\rm geom}}
\def\gtog{\langle\rho\chi\phi|e^a_1\eta_2\rangle}
\def\pd#1{{\delta\ \over\delta #1}}
\def\Pr{\Pi_\rho}
\def\Pp{\Pi_\phi}
\def\Pc{\Pi_\chi}
\def\kh{\hat\kappa}

% Titlepage macros
\hoffset=0.2truein
\voffset=0.1truein
\hsize=6truein
\def\TITLEPAGE{\frontpagetrue}

\def\BU{
        \address{{}\break
                Department of Physics\break
                Boston University\break
                590 Commonwealth Avenue\break
                Boston, Massachusets\ \ 02215\ \ \ U.{}S.{}A.}}

%%%%%

\def\TITLE#1{\vskip.5in \centerline{\fourteenpoint#1}}
\def\TITLECONT#1{\medskip\centerline{\fourteenpoint#1}}
\def\AUTHOR#1{\vskip.2in \centerline{#1}}
\def\ABSTRACT#1{\vskip1in \vfil \centerline
            {\twelvepoint \bf Abstract}
                     #1 \vfil}
\def\ENDTITLEPAGE{\vfil \eject \pageno=1}
\hfuzz=5pt
\tolerance=10000
\TITLEPAGE
\TITLE{A COMPARISON OF TWO QUANTIZATION PROCEDURES}
\TITLECONT{FOR LINEAL GRAVITY${}^*$}\footnote{*}
{This work was supported in part by funds provided by the U.{}S.
Department of Energy (D.{}O.{}E.) under contract \#DE-FG02-91ER40676}
\AUTHOR{Eric Benedict}
\BU
\ABSTRACT{
We consider two programs for quantizing gravity in $1+1$ dimensions,
which have appeared in the literature:
one using a gauge--theoretic approach and the other following a more
conventional ``geometric'' approach. We compare the wave functionals
produced by the two different programs by finding matrix elements
between the variables of the two theories. We find that the
wave functionals are equivalent.
}
\vfill
\hbox to \hsize{BU-HEP 94-15\hfill July 1994}
\ENDTITLEPAGE
\eject

%\chapter{Introduction}

Gravitational theories have been succesfully quantized in $1+1$
dimensions. These theories are not based on the Einstein equations,
which do not exist in $1+1$ dimensions.
Rather one considers an action of the form
$$I=\int d^2x \sqrt{-g} (\phi R-V(\phi))\eqno(1)$$
for particular choices of $V(\phi)$. We shall focus on the
string--inspired model where $V(\phi)=\Lambda$,
a model that has been quantized using different
methods. The first approach\Ref\Jac{
D.~Cangemi, in {\sl Proceedings of XXIIIth International Conference
on Differential Methods in Theoretical Physics}, Ixtapa, Mexico (in press);
R.~Jackiw in {\sl Proceedings of IMP 93}, Hangzhou, China (World
Scientific, Singapore, in press);
D.~Cangemi, R.~Jackiw, {\sl Phys. Rev. D}
(in press).}
\Ref\Amati{D.~Amati, S.~Elitzur, E.~Rabinovici, {\sl Nucl. Phys.}
{\bf B418} (1994) 45.}
uses a ``gauge formulation'':
\footnote{\dag}{T.~Strobl%
\Ref\Strobl{T.~Strobl, Technische Universit\"at Wien preprint,
TUW-93-26, 1993.}
followed a closely related approach, using a first--order geometric action.
}
beginning with an action invariant
under the extended Poincar\'e group, one shows that upon solving some
of the (gauge) constraints, the action reduces to (1). The second approach%
\Ref\Kun{D.~Louis--Martinez, J.~Gegenberg, G.~Kunstatter, {\sl Phys.
Lett.} {\bf B321} (1994) 193.}
uses a more conventional ``geometric formulation'' in which (1)
is written using a specific parametrization for the metric. This
procedure leads to a Hamiltonian that is a
sum of diffeomorphism constraints.
We shall show the relationship between the wave functionals obtained with
the two methods. The approach we
shall take is to
construct a canonical transformation between the variables in one
theory with those of the other, and use it to obtain
matrix elements between the variables of the two theories. As a
result of our investigation we shall find that the geometric
wave functional is equivalent to the gauge--theoretic wave functional.

We shall denote spacetime indices by lower case Greek letters, and
tangent space indices by lower case Latin letters. Timelike vectors
are taken to have a positive squared length, and we raise and lower
tangent--space indices with the metric
$h^{ab}={\rm diag}(1,-1)$.
The sign of the
totally antisymmetric symbol in two dimensions
is defined by $\epsilon^{01}=1$.
We shall use the notation
$A^\pm\equiv{1\over\sqrt{2}}(A^0\pm A^1)$ to denote light--cone components.

%\chapter{Lagrangian in the Gauge Formulation}

In the gauge formulation\refmark\Jac the Lagrange density in
canonical form is:
$$\Lga=\eta_a\dot e^a_1+\eta_2\dot\omega_1+\eta_3\dot a_1+
  e^a_0 G_a+\omega_0 G_2+a_0 G_3\eqno(2)$$
where the constraints
$$\eqalignno{
G_a&=\eta_a'+\epsilon_a^{\ b}\eta_b\omega_1+\eta_3
  \epsilon_{ab}e^b_1&(3a)\cr
G_2&=\eta_2'+\eta_a\epsilon^a_{\ b}e^b_1&(3b)\cr
G_3&=\eta_3'&(3c)\cr
}$$
obey the algebra of the extended Poincar\'e group.
In the absence of matter we can remove the dynamics of the field $a$
from the Lagrangian by solving the constraint $G_3$. We do this
by setting $\eta_3$ to be a constant, $\Lambda$.
The term involving $a$ is now a total derivative,
$\int d^2x\>(a_0\eta_3'+\eta_3\dot a_1)$, and we ignore it henceforth.
With this simplification we can write
the action following from (2) as
$$\int d^2x\>\Lga=
  \int d^2x\>\bigl(\eta_a\dot e^a_1+\eta_2\dot\omega+e^a_0 G_a+\omega_0 G_2
  \bigr)\eqno(4)$$
If we now interpret $e^a$ as a {\it Zweibein} 1-form and $\omega$
as a spin connection 1-form,
and use the equations of motion to eliminate $\eta_a$,
the action (4) is equivalent to\refmark\Jac
$$I=\int d^2x \sqrt{-g} (\phi R-\Lambda)\eqno(5)$$
where $2\phi=\eta_2$ and $R$ is the scalar curvature.
This is the starting point for the geometric formulation.
By parametrizing the metric as
$$ds^2=e^{2\rho}({\cal A}dt^2-(dx+{\cal B}dt)^2)\eqno(6)$$
the action (5) reads\refmark\Kun
$$I\equiv\int d^2x\>\Lge=
  \int d^2x\>\bigl(\Pr\dot\rho+\Pp\dot\phi-{\cal BP}-{\cal AE}\bigr)\eqno(7)$$
The variables ${\cal A}$ and ${\cal B}$ are Lagrange multipliers
enforcing the constraints
$${\cal P}=\rho'\Pi_\rho+\phi'\Pi_\phi-\Pi_\rho'\>,\eqno(8)$$
which generates spatial diffeomorphisms,
\footnote{\ddag}{By defining the variables
$\xi\equiv e^{-\rho}\Pr$, $\Pi_\xi\equiv -e^\rho$,
(8) can be cast in a more
familiar form: $${\cal P}=\xi'\Pi_\xi+\phi'\Pp$$}
and
$${\cal E}=2\phi''-2\phi'\rho'-
  {1\over 2}\Pi_\phi\Pi_\rho+e^{2\rho}\Lambda\>,\eqno(9)$$
which generates time translations.

%\chapter{Finding the Canonical Transformation}

Having presented the two Lagrangians (4) and (7),
our goal is to find an explicit canonical relationship between
the variables of the two theories.
We meet a difficulty, however, since
there are six variables in the gauge theory, while there are only four
in the geometric theory. We shall therefore introduce a new geometric
variable $\chi$ along with its conjugate momentum.
We propose the following six equations for the canonical transformation:
\halign{\hskip 1.75truein $#$\qquad\hfil&$#$\hfil&\qquad\qquad\hfill$#$\cr
 & & \cr
\rho={1\over 2}\ln (-2e^+_1 e^-_1)&\Pr=\eta_a e^a_1&(10a)\cr
 & & \cr
\phi={1\over 2}\eta_2&\Pp+2\chi'=-2\omega_1&(10b)\cr
 & & \cr
\chi=-{1\over 2}\ln(-e^+_1/e^-_1)&\Pc-2\phi'=
   \eta_a\epsilon^a_{\ b} e^b_1&(10c)\cr
 & & \cr
}\noindent
The Lagrange multipliers are also related:
we recover
the metric (6) in the gauge formulation if we write
$$e^a_0={\cal A}\epsilon^a_{\ b}e^b_1+{\cal B}e^a_1\eqno(11)$$
where $e^a_\mu$ is to be interpreted as a {\it Zweibein}.
Inserting this and the variables defined by (10a-c) into $\Lga$
in (4) we find that
$$\Lga=\Lge+\Pc\dot\chi+\Pc\omega_0-
  {\cal A}(\Pc\rho'-\Pc')+
  {1\over 2}{\cal B}\,\Pp\Pc-{d\ \over dt}(\Pp\phi)\eqno(12)$$
We then see that aside from the total time derivative (which will
produce a difference in phase between the two wave functionals) the
theories are equivalent if we first solve the constraint $\Pc=0$.
We recognize this constraint to coincide with $G_2$: both are
multiplied by $\omega_0$; moreover, from (10b,c) and (3b) we see that
$\Pc=G_2$.

We have demonstrated the equivalence of the gauge and geometric
formulations on the classical level. We now show the
relationship at the quantum level. In either approach,
we obtain the physical wave functional by demanding that the constraints
annihilate it.
In the gauge formulation we shall find it most convenient to write the
wave functional in the variables $(e^a_1,\eta_2)$, allowing
$(\eta_a,\omega_1)$ to act by differentiation. We shall obtain this wave
functional by Fourier transforming the solution written in terms of the
variables $(\eta_a,\eta_2)$, which has appeared in the
literature\refmark\Jac\refmark\Amati\refmark\Strobl.
This solution has the form
$$\langle\eta_a\eta_2|\Psi\rangle=
\delta(\eta_a\eta^a-2\Lambda\eta_2-M)
e^{iS(\eta_a,\eta_2)}
f(M)\eqno(13)
$$
with $M$ constant and $f$ an arbitrary function of $M$. The phase is
given by
$$S(\eta_a,\eta_2)
  \equiv\int\eta_2{\epsilon^{ab}\eta_a\eta_b'\over
  \eta_c\eta^c}\> dx\eqno(14)$$
We obtain the solution in terms of $(e^a_1,\eta_2)$ by performing
the functional integral
$$\langle e^a_1\eta_2|\Psi\rangle=
  \int{\cal D}\eta_a e^{i\int\eta_a e^a_1\>dx}
  \langle\eta_a\eta_2|\Psi\rangle
\eqno(15)
$$
We solve this integral by inserting the identity
$$1=\int{\cal D}\theta\>\delta\left({1\over 2}
   \ln{\eta_+\over\eta_-}-\theta\right)\eqno(16)$$
into (15). Upon defining the tangent--space vector
$$\kappa_a(\theta)
   \equiv\sqrt{M+2\Lambda\eta_2}(\cosh\theta,\sinh\theta)\eqno(17)$$
we find
$$\eqalignno{
\langle e^a_1\eta_2|\Psi\rangle
&=\int{\cal D}\theta\int{\cal D}\eta_a\>\delta\left({1\over 2}
   \ln{\eta_+\over\eta_-}-\theta\right)e^{i\int\eta_a e^a_1\>dx}
  \langle\eta_a\eta_2|\Psi\rangle&{}\cr
&={\cal N}\int{\cal D}\theta\>\exp\left\{i\int\kappa_a(\theta)e^a_1\>dx+
   iS(\kappa_a(\theta),\eta_2)\right\}f(M)&(18)\cr
}$$
where the normalization ${\cal N}$ absorbs field--independent multiplicative
factors. The integral (18) can be performed if we take it to be the
limit of a sequence of finite--dimensional integrals, where
space has been discretized. Absorbing constant factors into a
redefined normalization constant ${\cal N}'$ we can express the result
as a product of Bessel functions:
$$\eqalignno{
\int {\cal D}\theta\>\Psi(e^a_1&,\eta_2;\theta)&{}\cr
&={\cal N}'\lim_{N\rightarrow\infty}\prod_{i=1}^N
\left[\left({e^+_1\over e^-_1}\right)^{i\eta_2'\Delta x_i}
K_{i\eta_2'\Delta x_i}
\left(\Delta x_i
\sqrt{\strut -e^a_1 e_{a1}}
\sqrt{\strut\kappa_a\kappa^a}
\right)\right]_{x=x_i}&(19)
\cr}$$
In the continuum limit $\Delta x_i\rightarrow 0$
this expression diverges.
One can verify, however, that gauge--invariance holds to
${\cal O}(\Delta x_i)$ for every finite--dimensional integral.
One would like to find a way to regularize this functional integral
without destroying gauge--invariance. The author was unable to do this by
straightforward manipulation of (19). We shall therefore follow a more
indirect approach.

Consider the integrand of (18)
$$\Psi(e^a_1,\eta_2;\theta)={\cal N}\exp\left\{i\int\kappa_a(\theta)e^a_1\>dx+
  iS(\kappa(\theta),\eta_2)\right\}\eqno(20)$$
Using the expression (17) for $\kappa_a$ and (14) for
$S(\kappa_a,\eta_2)$, we can write this as
$$\Psi(e^a_1,\eta_2;\theta)={\cal N}\exp\left\{i\int dx\>\bigl(
  \kappa_a(\theta)e^a_1+\eta_2\theta'\bigr)\right\}\eqno(21)$$
We obtained this expression by inserting
$\delta({1\over 2}\ln(\eta_+/\eta_-)-\theta)$ into the integral (15),
that is, by fixing the ``angular'' dependence of $\eta_a$ within the
integral. This hyperbolic angle can be changed by a gauge transformation
generated by the constraint $G_2$. As a consequence we do not expect
(21) to be gauge--invariant for an arbitrary $\theta$. We may,
however, let $\theta$ depend upon $e^a_1$ and $\eta_2$ in such a way
that (21) is invariant. One can verify
by direct calculation using (21) that the condition that $G_2$
annihilates $\Psi(e^a_1,\eta_2;\theta)$,
$$\left(i e^a_1\epsilon_a^{\ b}\pd{e^b_1}+\eta_2'\right)
  \Psi(e^a_1,\eta_2;\theta)=0\eqno(22)$$
is equivalent to demanding that
$$\pd\theta\Psi(e^a_1,\eta_2;\theta)=0\eqno(23)$$
When the conditions (22) and (23) hold
one may check that the constraints $G_a$ also annihilate the
wave functional. [Note that (23) implies that $\theta$ can
be treated as independent of the other
variables when applying functional
derivatives to $\Psi(e^a_1,\eta_2;\theta)$.] The wave functional
(21), subject to the condition (23), is then the well--defined,
gauge--invariant solution that we sought.

%\chapter{The wave functional in the Geometric Formulation}

In the geometric formulation we
again obtain the physical wave functional by applying the constraints to it.
Rather than applying
${\cal P}$ and ${\cal E}$ (Eqs 8 and 9) directly, however, we form the linear
combination\refmark\Kun
$$-e^{-2\rho}(\phi'{\cal E}+{1\over 2}\Pi_\rho{\cal P})
=\left[
{e^{-2\rho}\over 4}\left(
\Pi_\rho^2-(2\phi')^2
\right)-\Lambda\phi
\right]'\eqno(24)
$$
Using the canonical transformation (10a-c), and setting $\Pc=0$,
one may verify that the quantity in brackets is just
$M/4$, where $M$ is the same invariant as in the gauge theory.
We therefore write the constraint (24) as
$$e^{-2\rho}\left(
\Pi_\rho^2-(2\phi')^2
\right)-4\Lambda\phi-M=0\eqno(25)
$$
Solving this expression for $\Pi_\rho$ and
solving ${\cal E}$ for $\Pi_\phi$
we apply the constraints in the following way:
$$\eqalignno{
\left(\Pi_\rho-Q\right)\Psi&=0&(26a)\cr
\left(\Pi_\phi-{g\over Q}\right)\Psi&=0&(26b)\cr
}$$
where
$Q\equiv \sqrt{(2\phi')^2+(4\Lambda\phi+M)e^{2\rho}}$
and
$g\equiv 4\phi''-4\phi'\rho'+2\Lambda e^{2\rho}$.
The solution is\refmark\Kun
$$\Psi_{\rm geom}(\rho,\phi)=
\exp\left\{
i\int dx\>\left(Q+\phi'\ln{2\phi'-Q\over 2\phi'+Q}\right)
\right\}f(M)\eqno(27)
$$
where again $f$ is arbitrary.

%\chapter{Relating the wave functionals}

We shall now use the canonical transformation (10a-c)
to obtain the matrix elements relating the variables of the two theories.
We promote the canonical variables to
operators and apply $\langle\rho\chi\phi|$
on the left and $|e^a_1\eta_2\rangle$ on the right. The resulting set
of equations has the solution
$$\gtog=\delta(2\phi-\eta_2)
  \delta^2\bigl(e^a_1\mp e^\rho\epsilon^{ab}\kh_b(\chi)\bigr)
  e^{i\int\chi\eta_2'\>dx}
\eqno(28)$$
where $\kh_a\equiv\kappa_a/\sqrt{\kappa_b\kappa^b}$.
Examining (21) and (28) we find that
transforming from the gauge to the geometric wave functional only involves
integrating over a product of $\delta$-functions, so
we find immediately that
$$\eqalignno{\Psi(\rho,\chi,\phi;\theta)
&=\int{\cal D}\eta_a{\cal D}\eta_2\gtog
   \Psi(e^a_1,\eta_2;\theta)&\cr
&={\cal N}\exp\left\{i
\int dx\>\biggl(\pm\kappa_a(\theta)\epsilon^{ab}\kh_b(\chi)e^\rho-
   2\phi\chi'+2\phi\theta'\biggr)
\right\}f(M)&(29)\cr
}$$
We claim that this is equivalent to (27).
The equivalence is not manifest, however, so we shall
use the constraints to rewrite the wave functional (29) so that $\theta$ and
$\chi$ do not appear explicitly. When $\Psi(e^a_1,\eta_2;\theta)$
satisfies (23) it is annihilated by the constraint
$$\eta_a\eta^a-2\Lambda\eta_2-M\approx 0\eqno(30)$$
We showed that this constraint has a counterpart in the geometric theory
(Eq 25) which we wrote as $\Pr-Q(\rho,\phi)\approx 0$. This condition is
therefore an identity for the wave functional (29) which we can use to
identify $Q$ in terms of $\theta$ and $\chi$. Using (26a)
we discover the equality
$$Q=\pm\kappa_a(\theta)\epsilon^{ab}\kh_b(\chi)e^\rho\eqno(31)$$
We showed that solving the constraint $G_2\approx 0$, which in the
geometric variables becomes $\Pc\approx 0$, is equivalent to the
condition (23). Using either condition we obtain the second equality
$$2\phi'=\mp\kappa_a(\theta)\kh^a(\chi)e^\rho\eqno(32)$$
Using (31) and (32) we can eliminate $\theta$ and
$\chi$ in
terms of $Q$ and $2\phi'$, if we form the combination
$$\phi'\ln{2\phi'-Q\over 2\phi'+Q}
  =\phi'\ln{\kh_a(\theta)(h^{ab}+\epsilon^{ab})\kh_b(\chi)\over
             \kh_c(\theta)(h^{cd}-\epsilon^{cd})\kh_d(\chi)}
=2\phi'(\chi-\theta)\eqno(33)$$
Inserting this into the wave functional (29),
we obtain
$$\langle\rho\chi\phi|\Psi_\kappa\rangle=
{\cal N}\exp\left\{
i\int dx\>\left( Q+\phi'\ln{2\phi'-Q\over 2\phi'+Q}\right)
\right\}f(M)\eqno(34)
$$
which is---apart from the normalizing factor ${\cal N}$---the same as (28).

We have shown that the two theories are canonically related at the
classical level and, moreover, the wave functionals that result upon
quantization are equivalent. It is interesting to
note that the presentation of the wave
functional in the gauge variables illuminates a puzzle in the geometric
theory\refmark\Kun. Field configurations which yield an imaginary phase
correspond to classically forbidden regions of configuration space. In
the geometric variables it is difficult to see why an imaginary
contribution to $\ln\bigl[(2\phi'-Q)/(2\phi'+Q)\bigr]$ should be classically
forbidden. From (33), however, we see that in that case the ``angle''
of the tangent--space vector $e^a_1$ or $\eta_a$ becomes complex,
which does not occur in the classical gauge theory.

%\chapter{Conclusions and comments}

\vskip 0.5in\noindent
{\bf Acknowledgements}
\bigskip\noindent
I am grateful to
R.~Jackiw for suggesting the problem and for many useful
discussions. I am thankful both to him and D.~Cangemi
for making extensive comments on the manuscript.

\vfill\eject
\refout

\bye